\documentclass[12pt]{iopart}
\usepackage{iopams}  
\usepackage{xcolor} 
\usepackage{graphicx}
\usepackage{hyperref} 
\usepackage{dirtytalk}

\usepackage{graphicx}
\usepackage{float}
\usepackage{placeins}  

\newcommand{\workingtitle}{The two conduction bands of monolayer CrSBr on Au}
\newcommand{\shortenedtitle}{CrSBr on Au}

\newcommand{\authorlist}{
\author{Yogal Prasad Ghimirey}
\address{Diamond Light Source Ltd, Harwell Science and Innovation Campus, Didcot, OX11 0DE, UK}
\address{Department of Physics, University of Warwick, Coventry, CV4 7AL, UK}

\author{Laxman Nagireddy}
\address{Department of Physics, University of Warwick, Coventry, CV4 7AL, UK}
\address{CY Cergy Paris Université, CEA, LIDYL, 91191 Gif-sur-Yvette, France}
\address{Université Paris-Saclay, CEA, LIDYL, 91191, Gif-sur-Yvette, France}

\author{Cephise Cacho}
\address{Diamond Light Source Ltd, Harwell Science and Innovation Campus, Didcot, OX11 0DE, UK}

\author{Neil R. Wilson}
\address{Department of Physics, University of Warwick, Coventry, CV4 7AL, UK}
\ead{Neil.Wilson@warwick.ac.uk}

\author{Matthew D. Watson}
\address{Diamond Light Source Ltd, Harwell Science and Innovation Campus, Didcot, OX11 0DE, UK}
\ead{matthew.watson@diamond.ac.uk}
}

\begin{document}

\title[\shortenedtitle]{\workingtitle} 

\authorlist

\newpage

\begin{abstract}
We report the electronic structure of monolayer CrSBr exfoliated onto mica template-stripped gold substrates. Angle-resolved photoemission spectroscopy reveals charge transfer from the substrate, populating the conduction band of monolayer CrSBr, accompanied by a pronounced reduction in the quasiparticle band gap. Furthermore, we observe two separate conduction bands that exhibit a splitting at the X point. This indicates a breaking of glide-mirror symmetry, which in the bulk or in a free-standing monolayer protects the band degeneracies at the Brillouin zone boundary. Our results demonstrate that ultraflat gold substrates do more than modify carrier densities and screening: they can lift symmetry-protected degeneracies and thus fundamentally reshape the band topology of 2D materials.

\end{abstract}


%

\vspace{2pc}
\noindent{{\it \today}}

\section{Introduction}

The discovery of magnetism in atomically thin materials has opened new opportunities to realize exotic spin phenomena and novel functionalities through the exploration and exploitation of dimensional confinement, reduced symmetry, interfacial effects, and the use of tuning parameters such as strain and electric field \cite{grubisic-cabo_roadmap_2025}. Among the growing class of two-dimensional (2D) magnets \cite{Zhang2024}, CrSBr has emerged as a focus of investigation due to its unique combination of easily tunable magnetic order, relatively high N\'{e}el temperature, highly anisotropic optical properties, and exceptional environmental stability \cite{ziebel_crsbr_2024,Klein2024MaterialsBeyond}. This has rapidly led to the incorporation of mono- and few-layers of CrSBr into imaginative 2D heterostructure and device geometries. For example, such devices have been used to demonstrate: twist-assisted anti-ferromagnetic \cite{chen_twist-assisted_2024}, and strain-programmable \cite{cenker_strain-programmable_2025}, magnetic tunnel junctions; local control of superconducting spin valves \cite{jo_local_2023}; magnetic memory in twisted CrSBr bilayers \cite{boix-constant_multistep_2024}; proximity effects at the graphene/CrSBr interface \cite{yang_electrostatically_2024,rizzo_engineering_2025}; and direct electrical detection of antiferromagnetic resonances for antiferromagnetic spintronics \cite{cham_spin-filter_2025}. 

Developing such electronic applications will require a detailed understanding of the electronic structure of CrSBr and of its behaviour when interfaced with metals. The interfaces formed between 2D materials and metals are of themselves an area of intense interest \cite{pirker_when_2024}: the strength of interaction can vary widely from weak van der Waals to strong covalent bonding, with charge redistribution or charge transfer, and in gap states that can result in Fermi level pinning. Strong interaction at the interface can break the symmetries expected for free-standing monolayer 2D materials, as proven by the observation of symmetry forbidden Raman modes in monolayer transition metal dichalcogenides \cite{rodriguez_activation_2022, Velick2018, Pollmann2021}. Investigating these interfaces is challenging: they are often heterogeneous because of rough polycrystalline metal surfaces \cite{Zhu2022,Boehm2023}, and the most commonly used characterisation techniques, such as Raman and photoluminescence spectroscopy, give only indirect information about the atomic and electronic structure. Angle-resolved photoemission spectroscopy (ARPES) has been widely used to directly interrogate the electronic structure of 2DMs, including CrSBr \cite{Bianchi2023ParamagneticSpectroscopy,Bianchi2023ChargeTransfer,Watson2024GiantExchange,wu_mott_2025,Biktagirov2025IntrinsicDefects,Smolenski2025LargeExciton}, and can probe the 2DM-metal interface due to its surface sensitivity \cite{Enderlein2010, Miwa2015, Brugger2009, cucchi_electronic_2021,Dreher2021, GrubisicCabo2023InMaterials, NagireddysAwesomeSaucePaper}. However, ARPES requires atomically flat surfaces. A well-controlled platform is therefore essential for isolating how substrate-induced perturbations reshape the low-energy electronic structure of monolayer CrSBr.

In this work, we investigate monolayer CrSBr on ultraflat template-stripped gold (TSG). Using ARPES, we find a substantial charge transfer from the substrate, with accompanying renormalisation of the band gap. A splitting of the conduction states into two separate bands is found not only at $\Gamma$, but also resolved at X, implying a breaking of the glide-mirror symmetry under which the two Cr sites in the nonsymmorphic unit cell are equivalent. The results show how the metal-2DM interface can have effects beyond simple charge transfer and band gap renormalisation; the interface can also split band degeneracies that would be symmetry-enforced in freestanding monolayers.

\section{Results}

\begin{figure}[t]   
\centering
     \includegraphics[width=1.0\linewidth]{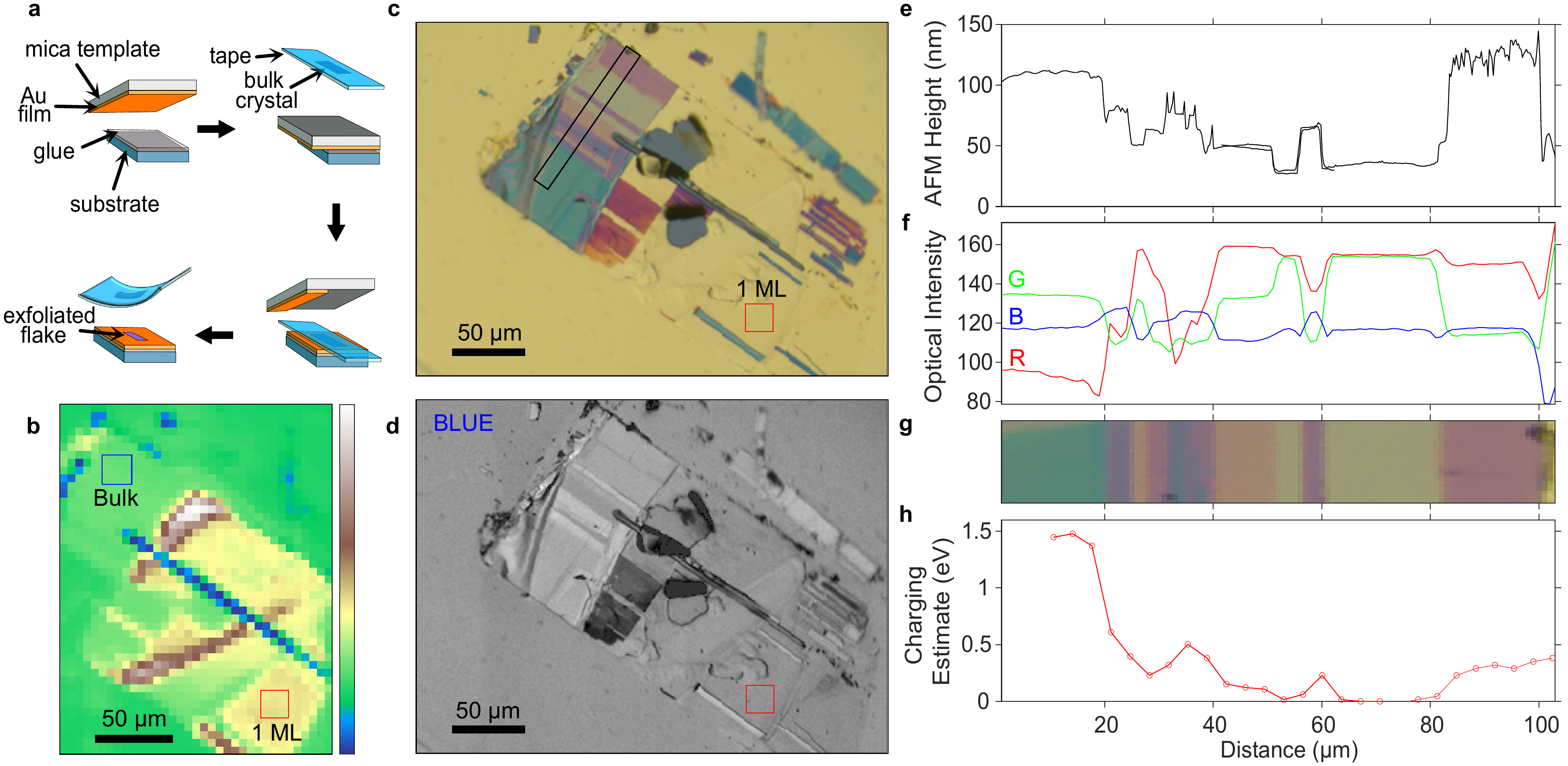}
\caption{\textbf{Characterisation of bulk-like and monolayer regions of CrSBr on TSG.} (a) Schematic of the TSG approach. (b) SPEM map, highlighting the 1 ML and bulk regions from which the data in Fig. 2 are measured. (c) Optical image. (d) Grayscale image of the blue channel only, which is the most sensitive to the thinnest regions. (e) AFM height profile, and (f) optical intensity within the region shown in (g), corresponding to the black rectangle in (c). (h) Estimate of the magnitude of sample charging during ARPES, as a function of position.}
    \label{fig1}
\end{figure}

Our approach to preparing thin flakes of CrSBr on Au utilises TSG \cite{Vogel2012AsProcedures}, as summarised in Fig.~\ref{fig1}(a). We explored many details of this method in another recent paper \cite{NagireddysAwesomeSaucePaper}, so we briefly recap the key ingredients here. It begins with the deposition of Au onto a freshly cleaved mica template, which is then glued onto a substrate. In an Argon-filled glove box, the template is removed, and a tape holding exfoliated flakes is pressed onto the freshly exposed Au surface. This assembly is transferred into the ultrahigh vacuum (UHV) system, where the tape is finally removed, exposing clean surfaces of the 2D material. ARPES can then be applied without any further sample pretreatment, using scanning photoemission microscopy (SPEM) to locate the 2D material flakes and identify suitable regions for study, as in Fig.~\ref{fig1} (b). The technique combines the principle of Au-assisted exfoliation of 2D materials \cite{Magda2015ExfoliationLayers,Huang2020UniversalCrystals} with a known route to prepare ultraflat gold films \cite{Vogel2012AsProcedures}, with the aim of minimizing surface roughness at the interface.

In the context of WSe\textsubscript{2} on (mica-templated) TSG, large regions of monolayer coverage are typically found \cite{NagireddysAwesomeSaucePaper}, but also few-layer and bulk-like regions. Our experience with CrSBr on TSG differs slightly in that we typically find either monolayer- or bulk-like (i.e. $\gtrsim 10$ nm) regions, but it seems to rarely yield measurable few-layer areas. Fig.~\ref{fig1}(c) shows an optical image of a typical CrSBr flake, prepared using the TSG method. The blue channel has the highest sensitivity to the thinnest parts of the flake, such as the area marked with a red box which we ascribe as having monolayer thickness. The optical contrast of monolayer CrSBr on TSG is not as clear as that of other 2D materials such as WSe\textsubscript{2}, but this region is clearly ultrathin. We further justify the ``monolayer" assignment in the Supplementary Figure S1, which is also supported by analysis of the core level spectra in Supplementary Figure S3. 

Compared with exfoliation onto SiO\textsubscript{2} \cite{Wilson2021InterlayerSemiconductor}, the CrSBr monolayer regions we obtain using TSG are much larger and more prevalent, consistent with the large monolayer regions obtained using similar Au-assisted exfoliation methods with other 2D materials \cite{Huang2020UniversalCrystals}. On the left side of the flake in Fig.~\ref{fig1}(c), the differently coloured rectangular shapes indicate bulk-like flakes of varying thickness. The zoomed image and optical intensity plots in Fig.~\ref{fig1}(f,g) show perfectly aligned steps, reflecting the orthorhombic crystal structure. We mapped the topography using atomic force microscopy, finding a range of flake thicknesses from 36 to 113 nm in Fig.~\ref{fig1}(e). 

The thicker parts of the flake show lower intensity in the spatial map of total counts (SPEM map), presented in Fig.~\ref{fig1}(b), for two reasons: firstly, the full attenuation of the relatively bright photoemission intensity from the underlying gold; and secondly, the CrSBr bands are shifted to higher binding energy due to sample charging, shifting the intensity out of the detection range of the analyser. In Fig.~\ref{fig1}(h) we quantify this charging effect, which appears to scale faster than linearly with sample thickness, and is close to 1.5 eV for the thickest region with 113 nm height (see also Supplementary Figure S2). We note that the magnitude of the sample charging effect is specific to the measurement temperature of 25 K and photon flux through the capillary mirror at the micro-ARPES beamline of I05 at Diamond Light Source, at a photon energy of 56 eV. However, it nicely illustrates how, at low temperatures, sample charging effects can be severely restrictive, as was found in previous measurements on bulk single crystals \cite{Bianchi2023ParamagneticSpectroscopy,Smolenski2025LargeExciton,Biktagirov2025IntrinsicDefects}. While there are multiple possible approaches to mitigating the problem of sample charging \cite{wu_mott_2025}, the analysis here emphasizes the usefulness of the TSG approach for obtaining low-temperature ARPES data on CrSBr, and 2D semiconducting samples more generally, with the non-charging regime established for flake thicknesses $\lesssim 30$~nm. 

\begin{figure}[t]   
\centering
        \includegraphics[width= 1.0\linewidth]{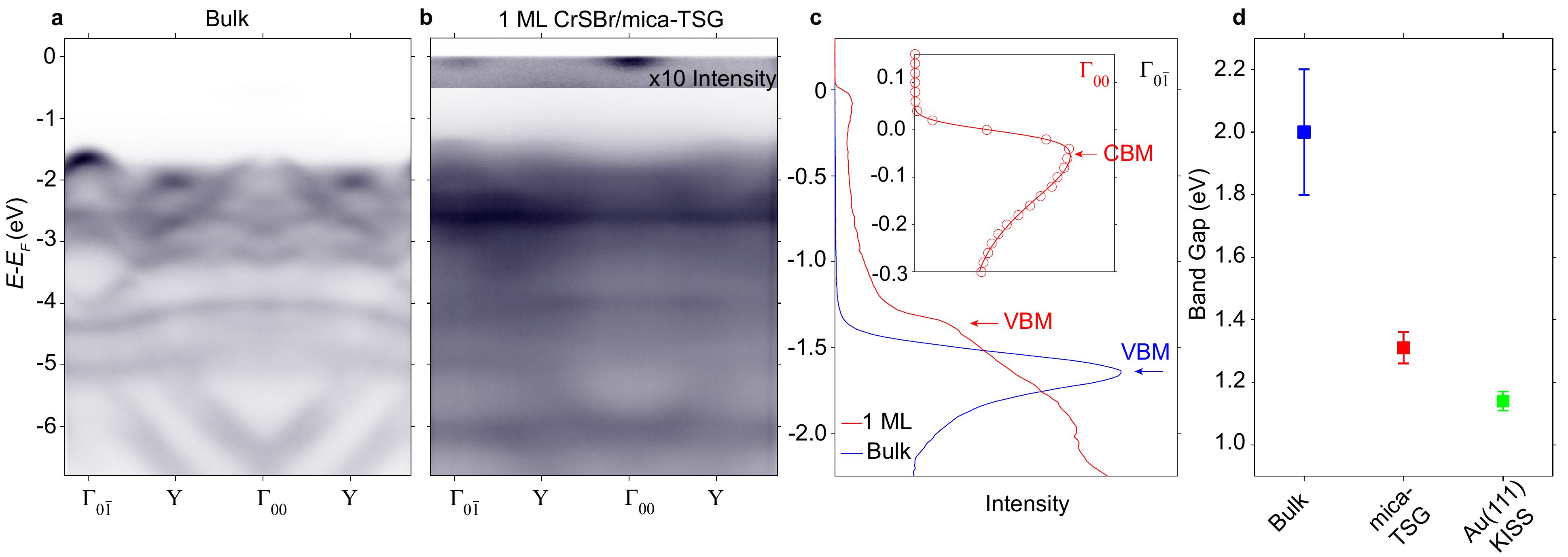}
    \caption{\textbf{Evolution of the band gap in the monolayer limit}. (a) ARPES spectra measured with $hv$=80 eV on the bulk-like region. The VBM is identified by a bright feature at the $\Gamma_{0\bar{1}}$ point in the second Brillouin zone. (b) Equivalent measurement on the monolayer region, highlighting the relatively weak spectral weight on the conduction band states. (c) EDCs from the $\Gamma_{0\bar{1}}$ points, with arrows showing the features associated with the VBMs in the bulk and monolayer cases. The inset shows a fit to the EDC at $\Gamma_{00}$, from which the CBM is extracted. (d) Comparison of the band gap of bulk and monolayer CrSBr on mica-TSG and also the Au(111) substrate via the KISS method reported by Bianchi \textit{et al.} \cite{Bianchi2023ChargeTransfer}.}
    \label{fig2}
\end{figure}

\begin{figure}[htbp]
    \centering
    \includegraphics[width=0.9\linewidth]{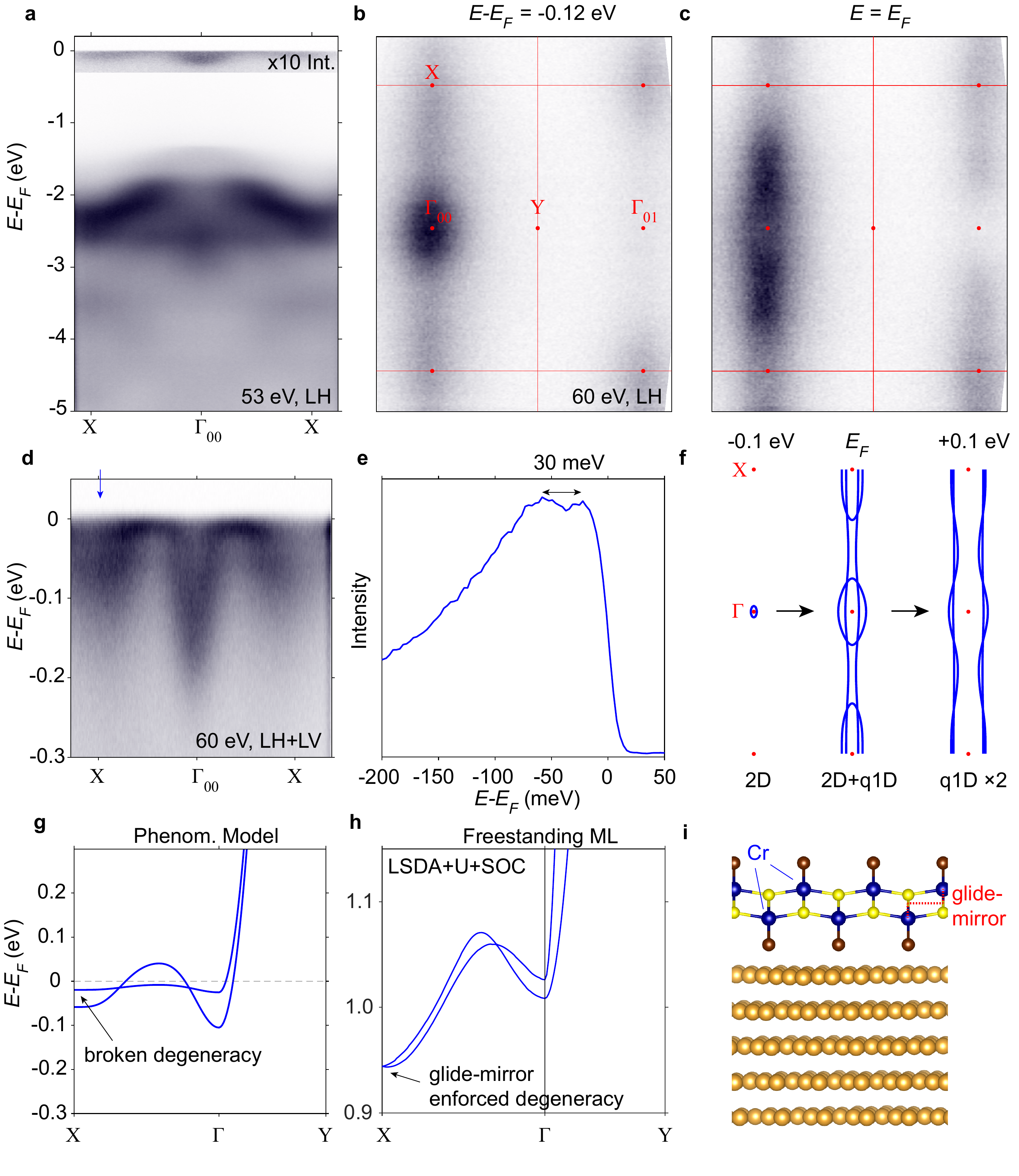}
    \caption{\textbf{Dimensionality crossover as a function of conduction band filling}. 
    (a) ARPES spectra on the 1 ML region along $\Gamma-\rm{X}$.
    (b) Constant energy contour at $-0.12$ eV, corresponding to the CBM of the lower conduction band.
    (c) Constant energy contour at $E_F$.
    (d) Dispersion along $\Gamma-\rm{X}$ with an EDC at $\rm{X}$ highlighting contributions from two electron bands.
    (e) EDC at $\rm{X}$ showing a splitting of 30 meV.
    (f) Schematic band structure evolution with conduction band filling.
    (g) Schematic band structures.
    (h) Density functional theory calculation of freestanding monolayer showing CB degeneracy at $\rm{X}$ enforced by glide-mirror symmetry.
    (i) Structural representation of monolayer CrSBr on Au(111). The two Cr sites would be equivalent in a freestanding monolayer via the glide-mirror symmetry, but become inequivalent due to the substrate.}
    \label{fig3}
\end{figure}

The electronic structure of the monolayer CrSBr on TSG differs substantially from the bulk, most notably in that charge transfer at the interface creates a significant population of the conduction band states, as seen in Fig.~\ref{fig2}(a,b). Since both the conduction and valence band edges are occupied, the band gap can be directly determined by analysis of the ARPES data. Fitting of the the energy dispersion curve (EDC) at $\Gamma_{00}$ yields a peak at -0.05 eV for the conduction band minimum (CBM) - though in fact there are two bands at $\Gamma$, only resolved in another measurement geometry (see later Fig.~\ref{fig3}), which means we should associate this estimate with a large uncertainty; we take -0.05(5) eV. The valence band maximum (VBM) is more subtle but manifests as a shoulder in the EDC at $\Gamma_{0\bar{1}}$ \cite{Bianchi2023ParamagneticSpectroscopy} and from a second derivative analysis we identify the VBM at -1.36(2) eV. The band gap is thus estimated at 1.31(6) eV for monolayer CrSBr on TSG. 

To make the comparison with the band gap in the bulk case, we must first acknowledge that there is some controversy regarding the magnitude of the fundamental band gap in CrSBr \cite{Klein2024MaterialsBeyond}. In our bulk data in Fig.~\ref{fig2}(a), at a temperature of 25~K, deep in the antiferromagnetic ground state, the VBM is at -1.64 eV but the chemical potential lies in the band gap and there is no population of the conduction band states. As argued previously, \cite{Bianchi2023ParamagneticSpectroscopy,Watson2024GiantExchange}, this implies a band gap well in excess of 1.64 eV, but the bulk band gap cannot be directly determined by ARPES alone. While experimental data from tunnelling and some earlier interpretation of optics have been interpreted as supporting a band gap around 1.5 eV, QSGW calculations coalesce around a band gap of approximately 2 eV \cite{Watson2024GiantExchange,Smolenski2025LargeExciton}, which seems to agree with the optical spectroscopy results with light polarised along the $a$ axis \cite{Shao2025MagneticallyConfined} and is more compatible with the ARPES data \cite{Watson2024GiantExchange,Bianchi2023ParamagneticSpectroscopy,Smolenski2025LargeExciton}.

Assuming this value of 2.0 eV for the bulk band gap, the 1.31 eV band gap measured here for the monolayer on TSG indicates a large band gap renormalization. For comparison, the effect is larger than the results in a gated monolayer WSe\textsubscript{2} \cite{Nguyen2019}, where the injection of electrons caused at most a renormalization from $\sim$2 to $\sim$1.7 eV. However, a smaller gap of 1.14(3) eV was reported for CrSBr/Au(111) prepared by the KISS technique \cite{Bianchi2023ChargeTransfer}, which also has a deeper CBM implying a larger charge transfer. A monotonic relationship between the induced charge carrier density and the renormalization of the band gap was also found by K dosing of the surface \cite{Smolenski2025LargeExciton}, and a gap of 1.3 eV was found after Rb dosing \cite{wu_mott_2025}, although we caution against quantitative comparisons due to differences in the methods used to estimate the band gap in each case. 

As an aside, we note that our bulk samples are truly semiconducting, not degenerately $n$-doped. Since we have the Fermi level reference from the gold substrate, and furthermore since charging effects are mitigated, we can confidently state that there is no intensity at $E_F$ in the bulk. However, in both bulk and monolayer data we find, with variable intensity, an in-gap state (Supplementary Figures S7-S9). The intensity is peaked at $E-E_F \sim -0.6$~eV, and distributed in momentum space, consistent with localised vacancies or other point-like defects. The most common defect in CrSBr is reported to be Br vacancies \cite{klein_sensing_2023}, and calculations including single Br vacancies yield a well-separated and flat in-gap state \cite{klein_sensing_2023,weile_defect_2025} that is qualitatively consistent with our data. While the influence of crystal growth methods on the defect types and density is starting to be understood \cite{ziebel_crsbr_2024}, in the context of ARPES measurements there are additional considerations of exposure to vacuum and the beam. Anecdotally, we have found that the photon beam can cause some time-dependent effects. These can be avoided by choosing lower photon energies, minimizing unnecessary beam exposure, and occasionally slightly moving the sample to avoid cumulative damage. 
Note that a previous report of ARPES on bulk samples claimed occupation of the conduction band due to defects intrinsic to CrSBr \cite{Biktagirov2025IntrinsicDefects}. We can confidently exclude such effects from our results: as stated above, there is no intensity at $E_F$ in the bulk and flakes thicker than $\gtrsim 50$ nm charge under the beam and hence are clearly insulating.

The population of the conduction bands in the monolayer on TSG gives us experimental access to their dimensionality. Since the sample is a single layer, we would not expect any $k_z$-dependence, and indeed our photon energy-dependent measurements (Supplementary Figure S6) show no out-of-plane dispersion in the valence bands. A photon energy of 53 eV, used in Fig.~\ref{fig3}(a) shows the valence bands most clearly, however the measured intensity of the conduction band is strongly peaked at photon energies around 60 eV, which we use in the remainder of Fig.~\ref{fig3} for high-resolution studies with the sample aligned in the $\Gamma-\rm{X}$ geometry. While the conduction bands are understood to be highly dispersive along $\Gamma{}-\rm{Y}$, arising from their particularly anisotropic orbital character \cite{Smolenski2025LargeExciton}, the perpendicular dispersion along $\Gamma{}-\rm{X}$ is a subtle question. The spectral weight at the conduction band edge in Fig.~\ref{fig3}(b) is centered at the $\Gamma$ point, while the dispersion in Fig.~\ref{fig3}(d) allows us to clearly see that the conduction band minimum is at $\Gamma$, although a local minimum also exists at X. In contrast, Smolenski \textit{et al.} placed the CBM at X ~\cite{Smolenski2025LargeExciton}, at least for their highly doped samples, while Bianchi \textit{et al.} found it at $\Gamma$ for CrSBr/Au(111) but at X for CrSBr/Ag(111) \cite{Bianchi2023ChargeTransfer}. It may be that the location of the CBM is dependent on the filling of the conduction band; similar phenomenology has been observed, for example, in alkali-dosed MoSe\textsubscript{2} \cite{Jung2024HolsteinPolarons}. Since we have a smaller charge transfer in our case than in Refs.~\cite{Bianchi2023ChargeTransfer,Smolenski2025LargeExciton}, our result of a CBM at $\Gamma$ (and by implication a direct band gap) is more likely to be representative of charge neutral semiconducting monolayers, and also the bulk.

An important consideration is that in a monolayer of CrSBr there should theoretically be two fully spin-polarised conduction bands, and here we find evidence for both states. From the $\Gamma$ point, there is a relatively steep upward dispersion, with a trajectory that appears to be heading above $E_F$. Separately to this, however, there also appears an almost flat band just below $E_F$, most prominent in between $\Gamma$ and X. This is best understood if there are two bands, and in Fig.~\ref{fig1}(h) we present a two-band scenario with a phenomenological model that is closely based on the data (Supplementary Figure S4). The most direct evidence for the presence of two bands is the observation of two distinct peaks in the EDC at the X point in Fig.~\ref{fig3}(e), although their separation is comparable to the spectral broadening from disorder. 

The two-band scenario implies a complex answer to the question of the dimensionality of the conduction bands, shown schematically in Fig.~\ref{fig3}(f). For a band filling just above the CBM, the constant energy contour contains a single 2D pocket around $\Gamma$. However, at the experimental Fermi level, the Fermi surface shown in Fig.~\ref{fig3}(c) has contributions from both quasi-1D and quasi-2D bands. From evaluating the Luttinger count in the phenomological model, we estimate that the experimental band filling corresponds to 0.05 $e^-$ per Cr (see Supplementary Figure S4). Only when both bands are fully occupied along the entire $\Gamma-\rm{X}$ dispersion, at higher filling, can one fairly describe the conduction bands as quasi-1D. 

The observation of the splitting of the two bands, represented in Fig.~\ref{fig3}(g), is particularly intriguing, since the band splitting at the X point should be forbidden in the nonsymmorphic space group $Pmmn$ due to glide-mirror symmetry \cite{Watson2024GiantExchange}. Illustrating this, Fig.~\ref{fig3}(h) shows a density functional theory (DFT) calculation for a free-standing monolayer of CrSBr, which includes a degeneracy of the conduction band states at the X point (see Methods and Supplementary Figure 5 for more details). It is worth remembering that a ``monolayer" of CrSBr actually contains Cr atoms at two different heights within the unit cell, which are equivalent under the intra-unit cell $n$ glide-mirror operation. However, as depicted in Fig.~\ref{fig3}(i), the presence of the Au substrate breaks this symmetry and breaks the equivalence of the Cr atoms. We hypothesise, therefore, that the splitting of the symmetry-enforced degeneracy can be ascribed to anisotropic interaction with the substrate.

\section{Discussion and conclusions}

Not all monolayers of 2D semiconductors show a population of the conduction band from charge transfer when prepared using TSG. In the prototypical case of monolayer WSe\textsubscript{2}, no hint of the conduction band was found when prepared with mica-TSG \cite{NagireddysAwesomeSaucePaper}. Similarly, using the KISS method, there is no population of the conduction band of WSe\textsubscript{2} \cite{GrubisicCabo2023InMaterials}, nor is there for MoS\textsubscript{2} grown on Au(111) \cite{Bruix2016Single-layerInteraction}. This cannot be easily reconciled by simple considerations of the work functions of the respective materials, and this suggests an important role for interfacial chemistry in determining the degree of charge transfer. 

The comparison with CrSBr prepared using the KISS technique \cite{Bianchi2023ChargeTransfer} is instructive. In the case of the KISS approach, there is a clean Au(111) surface prepared by cycles of sputtering and annealing in UHV, and a large charge transfer was found: the CBM is at -187 meV \cite{Bianchi2023ChargeTransfer}, compared with -50 meV for the average of the two conduction bands we find here. Most likely the difference arises because, although our Au is ultraflat, it is not ultraclean: the mica is briefly exposed to a cleanroom atmosphere before depositing Au, and after the mica is cleaved away, the TSG is briefly exposed to the atmosphere of the glove box (see Methods). In this sense, the TSG approach is closer to device fabrication scenarios, and our results imply that a population of the conduction bands may occur at CrSBr-Au interfaces, even if the Au is not ultraclean.  

The filling of the two conduction bands in our data places the Fermi level very close to the band crossings along the $\Gamma-\rm{X}$ direction. This places CrSBr/TSG intriguingly close to the regime where non-linear optical and transport properties are proposed to be topologically protected and governed by the quantum metric \cite{das_surface-dominated_2025}. However, the experimental data of monolayer CrSBr on Au points to two crossings along $\Gamma-\rm{X}$ rather than one as predicted by calculations, and the bands are found to have a 30 meV splitting at $\rm{X}$ where a free-standing monolayer should have strict band degeneracy. This raises the question of whether the principle of quantum metric governance survives the substrate-induced perturbation that breaks the glide-mirror symmetry. 

Substrate-induced symmetry-lowering is a quite general effect that can take many forms \cite{Bruix2016Single-layerInteraction}. Our results bring into focus the specific case of 2D materials with $a,b$ or $n$ glide-mirrors, i.e. intra-unit cell in-plane translation combined with $z\rightarrow-z$. These non-symmorphic symmetry elements generically protect band crossings at the Brillouin zone boundaries for freestanding monolayers, but our data exemplifies how these can be split by the symmetry-breaking effect of the substrate. 

To conclude, our results demonstrate that the interaction between monolayer CrSBr and an atomically flat gold substrate induces significant electronic modifications. Charge transfer from the substrate populates the conduction band and renormalizes the band gap. We observe two separate conduction bands at both the $\Gamma$ and X points. The latter indicates a breaking of the glide-mirror symmetry that renders the two Cr sites equivalent in a freestanding monolayer. These findings underscore that substrate coupling — especially to templated, ultraflat metals — can profoundly alter the symmetry and electronic structure of 2D materials, with implications for both fundamental studies and device integration.

\section*{Methods}
\noindent{}\textbf{Sample preparation:} \newline
The first step of sample preparation involved depositing gold using a Moorfield nanoPVD-S10A magnetron sputtering system onto freshly cleaved mica templates (AGG250-2, Agar Scientific). The mica sheets were cleaved with a razor blade in a cleanroom environment just before placing in the sputtering system. For the mica template, gold was sputtered at a rate of approximately 0.57\,nm/s to a final thickness of $\sim$100\,nm. Following deposition, the gold-coated surface was glued to a silicon carrier wafer using Opti-tec 5054-1 two-part epoxy (Intertronics) and cured in ambient atmosphere at 150\,$^\circ$C for 10 minutes. A pristine gold surface was then exposed by cleaving away the mica template inside an argon-filled glovebox using a sharp razor blade. Within $\sim$1 minute of revealing the fresh gold surface, exfoliated flakes of the bulk CrSBr (prepared on Kapton tape) were pressed onto the prepared gold surface. The entire assembly, including the Kapton tape, was subsequently heated on a hot plate at 140\,$^\circ$C for one minute to enhance adhesion. Exfoliation was completed by removing the Kapton tape in one continuous action under ultra-high vacuum (UHV) conditions ($\sim$10$^{-8}$\,mbar), leaving thin flakes adhered to the ultra-flat template-stripped gold (TSG) surface. The sample assembly was immediately transferred to the analysis chamber.
CrSBr crystals were synthesised by HQ Graphene (Netherlands) using a chemical vapour transport growth method. \newline

\noindent{}\textbf{\textmu ARPES:} \newline
Experiments were performed at the I05 beamline, using the capillary mirror optic available on the nano-ARPES branch to achieve a beam spot of 4 \textmu{}m (full width at half maximum). The sample temperature was $\sim$25~K. The energy resolution was typically $\sim 50$ meV, except for the higher resolution measurements in Fig.~3 where it was 15-20 meV. The polarisation was linear horizontal (LH) except in Fig.~\ref{fig3}(d,e) where a sum of data from LH and linear vertical (LV) polarisations is presented. The photon energies are stated in the figure panels or the captions. Both bulk and monolayer-like spectra were reproduced at several positions of several samples. We adopt the notation of Bianchi \textit{et. al.} \cite{Bianchi2023ChargeTransfer} for the labeling of high symmetry points in the first and second Brillouin zones. \newline

\noindent{}\textbf{Atomic Force Microscopy:} \newline
Images were acquired with a Bruker Dimension Icon operated in Peak Force tapping mode.\newline

\noindent{}\textbf{DFT:} \newline
The DFT calculations shown in Fig. 3 were performed within Wien2k. The LSDA+U (local spin density approximation) approximation was used, including spin-orbit coupling. The U parameter was chosen to match the calculations in Bianchi \textit{et. al.} \cite{Bianchi2023ChargeTransfer}, though here we consider a monolayer rather than a bilayer. The full band structure including the valence bands is presented in Supplementary Figure S5. \newline

\section*{Data availability statement}
Data supporting the plots in the manuscript are available from the corresponding authors on a reasonable request.

\section*{Acknowledgements}
We thank R. Rong, R. Davitt, and A. Louat for technical support. We thank Diamond Light Source for access to beamline I05 under proposal numbers SI37658, SI36215 and SI39549 that contributed to the results presented here. \textit{Funding:} We acknowledge support from EPSRC grant EP/T027207/1. LN was supported by a EUTOPIA PhD Co-tutelle Programme. 

\section*{Author Contributions}
N.W., M.W., and C.C. conceptualised and supervised the project. Y.P.G. and L.N. performed the sample preparation. All authors participated in the measurements at the beamline. The AFM data was measured by Y.P.G. and N.W., and M.W. performed the DFT calculations. Y.P.G. led the data analysis and preparation of the figures. The text was written by M.D.W., Y.P.G., and N.W. with input and review from all authors.

\section*{Competing Interests}
The authors declare no competing interests.

\section*{Correspondence}
Corresponding authors: \newline{}
M.W. (matthew.watson@diamond.ac.uk) \newline{}
N.W. (Neil.Wilson@warwick.ac.uk)

\section*{References}

\begin{flushleft}


\end{flushleft}


\end{document}